\documentclass{article}

\usepackage{arxiv}

\usepackage[utf8]{inputenc} 
\usepackage[T1]{fontenc}    
\usepackage{hyperref}       
\usepackage{url}            
\usepackage{booktabs}       
\usepackage{amsfonts}       
\usepackage{nicefrac}       
\usepackage{microtype}      
\usepackage{lipsum}
\usepackage{amsmath}
\usepackage{graphicx}

\title{Non-rigid Registration Method between 3D CT Liver Data and 2D Ultrasonic Images based on Demons Model}

\author{
  Shuo Huang \\
 $ ^{a}$Shanghai United-imaging Healthcare Co., Ltd \\
  Jiading District, Shanghai, 201807, China\\
  $^{b}$School of Biological Sciences \& Medical Engineering\\
  Southeast University\\
  Sipailou No.2, Xuanwu District\\
  Nanjing, Jiangsu, 210096, China\\
  \texttt{h.s.3691831@163.com} \\
  \And
 Ke Wu, Xiaolin Meng and Cheng Li \\
  Shanghai United-imaging Healthcare Co., Ltd \\
  Jiading District, Shanghai, 201807, China\\
  \texttt{\{ke.wu, xiaolin.meng, cheng.li\}@united-imaging.com}
}

\begin{document}
\maketitle

\begin{abstract}
The non-rigid registration between CT data and ultrasonic images of liver can facilitate the diagnosis and treatment, which has been widely studied in recent years. To improve the registration accuracy of the Demons model on the non-rigid registration between 3D CT liver data and 2D ultrasonic images, a novel boundary extraction and enhancement method based on radial directional local intuitionistic fuzzy entropy in the polar coordinates has been put forward, and a new registration workflow has been provided. Experiments show that our method can acquire high-accuracy registration results. Experiments also show that the accuracy of the results of our method is higher than that of the original Demons method and the Demons method using simulated ultrasonic image by Field II. The operation time of our registration workflow is about 30 seconds, and it can be used in the surgery.
\end{abstract}


\section{Introduction}
Ultrasonography, as a real-time imaging technique with no radiation to both patients and doctors, has been widely studied for using in computer-assisted surgeries \cite{Ref1} \cite{Ref2}. The registration and fusion between ultrasonic image (US image) and computed tomography (CT) data is receiving a lot of attention in recent years, since it is hard for the doctors to locate the exact location only by the 2D ultrasonic image. Moreover, this technique also helps doctors to acquire more information from data of other modalities like CT or magnetic resonance imaging (MRI), which also helps to improve the outcome of diagnosis and treatment plan.
The aim of image registration is to obtain an optimal spatial transformation (deformation model) that can maximize the similarity measure (objective function) of the two images, as shown in Equation (1), where S and T are the similarity measure and the spatial transformation, respectively.
\begin{equation}
S\left(T\right)=S\left(A\left(\vec{X}\right),B\left(T\left(\vec{X}\right)\right)\right)
\end{equation}
Since image registration can be regarded as a multiple parameter optimization problem for transformation model parameters, as shown in Equation (2), some optimization methods are employed in the iterative process to obtain the optimal value of parameters of the spatial transformation T in Equation (1).
\begin{equation}
\hat{T}=\mathop{argmax}\limits_{T}{S\left(T\right)}\ or\ \hat{T}=\mathop{argmin}\limits_{T}{S\left(T\right)}
\end{equation}
The image registration methods can be divided into two main categories: rigid and non-rigid registration. The rigid transform can correct the rotation and translation of images, and it is often used as the initialization of the non-rigid transform \cite{Ref3}. The affine transform is a basic method of non-rigid registration, which can further correct the scaling and shear mapping of images \cite{Ref4}.

For the non-linear deformation, a lot of deformable registration methods with different deformation models, objective functions and optimization methods have been put forward \cite{Ref5} \cite{Ref6}, \cite{Ref7}. The Demons model based on the mutual information (MI) force and Quasi-Newton method is widely employed in image registration and the correction of the non-rigid deformation in different images in recent years. The Demons model is a representative optical flow model in the non-rigid registration of images, which has strong ability in dealing with the non-smooth deformation and has good anti-noise performance. The mutual information (MI) is widely employed in multi-modal registration algorithms. It has a property of generality and is efficient and reliable in multi-modal registration. And the Quasi-Newton methods are widely used optimization methods in the registration methods, which aims to accumulate information from the previous iterations and take advantage of it in order to achieve better convergence.

However, the performance of the Demons model on the registration between 3D CT liver data and 2D ultrasonic images needs to be improved. Due to the low signal-to-noise ratio and contrast of the ultrasonic image, the reflection of boundaries of organs and tissues and the shadowing of bones, the registration between ultrasonic images to CT data is challenging \cite{Ref8}. Especially when it comes to the Demons method based on the mutual information force, these differences severely influences the mutual information between images, which leads to the wrong registration results. Moreover, it is also difficult to acquire the correspondence CT slice directly by rigid registration with ultrasonic image, since the iterative process is easily to be trapped in the local optimum as a result that the spatial structure information of the ultrasonic image is limited. 

To overcome these challenges, a lot of methods has been put forward in these years \cite{Ref8} \cite{Ref9}. Typically, these approaches contains two steps. The first one is matching the world coordinate systems between CT volume data and ultrasonic image before the registration, which is called “calibration”. Matching the world coordinate systems is an important and efficient pre-processing method since setting a precise initial value is efficient in avoiding the local optimum and reducing the calculation time. The coordinates of the ultrasonic probe can be obtained by ultrasonic probe tracking devices \cite{Ref10}\cite{Ref11}\cite{Ref12}\cite{Ref13}\cite{Ref14}. After the extraction of the coordinates of probe, a lot of methods have been provided to obtain the transformation between the world coordinate systems and the coordinates of CT volume data or ultrasonic probe \cite{Ref13}\cite{Ref14}\cite{Ref15}. 

The second one is reducing the influence of differences between 3D CT or MRI volume data and ultrasonic images caused by the reflection and shadowing and improve the accuracy of registration. One research direction is to extract some control points from the data and images, register these points and then deform the data and images using the deformation field of these points \cite{Ref16}\cite{Ref17}\cite{Ref18}. Another research direction is to register the ultrasonic data with simulated ultrasonic data from the CT volume data or MRI volume data. This kind of methods can acquire better overall registration performance since more data are involved in the registration, especially in the parts that do not have control point. 

In recent years, many efficient simulation method has been put forward and employed to register the ultrasonic data and CT or MRI data \cite{Ref19}\cite{Ref20}. Field II is an efficient model-based simulation method, which describes the wave propagation using a spherical wave theory and incorporates the transducer effect by combining the impulse responses of points over the transducer surface \cite{Ref21}\cite{Ref22}\cite{Ref23}\cite{Ref24}. The accuracy of this numerical model relies on its discrete precision \cite{Ref21}. W. Wein et al have put forward a ultrasonic image simulation method through the simulation of large-scale reflection and transmission of sound waves and the intensity mapping of different tissue \cite{Ref20}\cite{Ref25}. These simulation method helps to reduce the differences between the simulated and the real ultrasonic data, thus can improve the registration accuracy between CT volume data and ultrasonic images. However, these methods all require a lot of calculation, and the calculation time needs to be reduced.

In this work, with a novel boundary extraction and enhancement method based on radial directional local intuitionistic fuzzy entropy in the polar coordinates, the non-rigid Demons model based on the mutual information force and Quasi-Newton method has been improved to improve the registration accuracy between the 3D CT volume data and the 2D ultrasonic images after the matching of world coordinate system. A new registration workflow has also been put up. Experiments on 3 ultrasonic images and 1 CT volume data show that our method can acquire high-accuracy registration results. Experiments also show that the accuracy of the results of our method is higher than that of the original Demons method and the Demons method using the simulated ultrasonic image by Field II.

\section{Methods}
In order to get good register results, a novel 2D US image to 3D CT volume registration method based on the non-rigid Demons method is put forward in this paper. To improve the performance of the Demons method, the Block Matching and 3D Filtering (BM3D) method \cite{Ref26}\cite{Ref27}\cite{Ref28} and Block Matching and 4D Filtering (BM4D) method \cite{Ref29} have been used to remove the additive random noise in US and CT data, respectively, and a novel ultrasonic image-like edge extraction and enhancement method for CT slice based on radial directional local intuitionistic fuzzy entropy in the polar coordinates has been put forward. 
\begin{figure}
  \centering
     \includegraphics[width=0.665\linewidth]{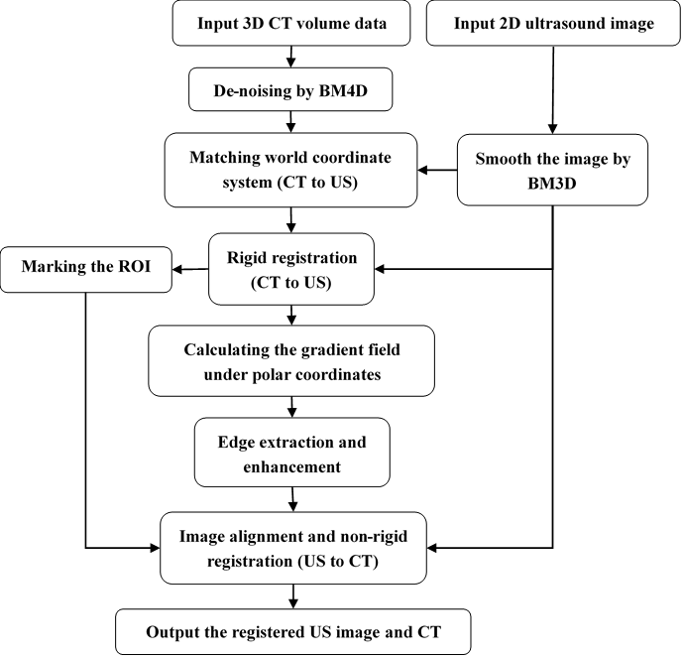}
     \caption{The flowchart of our method.}
  \label{fig:fig1}
\end{figure}

Our method contains three steps: the extraction of the corresponding CT slice, the edge enhancement of the CT slice and the non-rigid registration and resampling of the de-noised US image. The flowchart of our method is shown in Fig. 1.

\subsection{Data denoising}
The mutual information is employed as the similarity measure in the image rigid and non-rigid registration in this work. According to the definition of mutual information in Section Appendix A, suppressing the random noise (mostly the speckle and Gaussian noise) in the CT data and US images is an efficient way to improve the performance of the registration method, since such randomly deformed noise can severely influence the mutual information, and hereby reduce the accuracy of registration. 

The noise suppression technique has been widely studied in a long time period, and a lot of efficient approaches has been put forward \cite{Ref30}\cite{Ref31}\cite{Ref32}\cite{Ref33}\cite{Ref34}. The BM3D algorithm can acquire good de-noising effect with little influence on the details of images \cite{Ref30} when removing the additive noise such as the Gaussian noise and speckle noise \cite{Ref31}, which is multiplicative noise in definition but can be regarded as zero-mean additive noise according to the paper by J. S. Lee \cite{Ref34}. Moreover, on comparison with the denoising algorithms based on deep learning algorithms, the BM3D method is still competitive in the denoising performance. Therefore, the BM3D and BM4D methods are employed to remove the random noise in the US image and the CT data, respectively. In Section Appendix B, we briefly introduced BM3D and BM4D methods.

\subsection{The extraction of the corresponding CT slice}
To extract the corresponding CT slice of the US image, firstly, the world coordinate systems of the CT and ultrasound image are matched to align the data roughly. However, since the shooting of CT and US is not coinstantaneous, the slight movement of patient is ineluctable. To correct this movement and extract the corresponding CT slice of US image, the rigid registration algorithm is employed. 31 slices of the CT volume are extracted and de-noised by the BM4D method, where the roughly aligned corresponding slice is the 16th slice in the extracted volume. And the US image is de-noised by the BM3D method. Then the rigid registration algorithm is implemented on the de-noised CT and US data and the corresponding CT slice is extracted.

In this section, the affine transformation is employed to correct the rigid deformation \cite{Ref35}\cite{Ref36}\cite{Ref37}. The general affine model can describe the scale, rotation, shear and translation between two images. The transformation between pixel with coordinate $\left(x,y,z\right)$ in the fixed image and its corresponding position $\left(x\prime,y\prime,z\prime\right)$ in moving image is shown in Equation (3).

\begin{equation}
\left[ \begin{array}{c}x^\prime\\y^\prime\\z^\prime\end{array}\right]=\left[\begin{array}{ccc}q_{11}&q_{12}&q_{13}\\q_{21}&q_{22}&q_{23}\\q_{31}&q_{32}&q_{33}\end{array}\right]\left[\begin{array}{c}x\\y\\z\end{array}\right]+\left[\begin{array}{c}t_x\\t_y\\t_z\end{array}\right]
\end{equation}

Since image registration can be regarded as a multiple parameter optimization problem for transformation model parameters, as shown in Equation (2), the steepest decent method is employed in the iterative process to obtain the optimal value of parameters in Equation (3).

\subsection{Edge extraction and enhancement}
Another strategy to improve the performance of the 2D-2D Demons registration method is to enhance the edges such as the boundaries of the liver and the blood vessels in the CT image to improve the similarity between CT and US images. As shown in Chapter 1, this strategy has been widely used in the registration of CT and US data, and several efficient US image simulation approaches has been put forward. However, all these strategies need complicated calculation. As in the de-noising step, the scattering spots has been suppressed, high similarity in the smooth regions between de-noised CT and US images has been reached, a simple edge enhancement method is enough for acquiring high registration accuracy in the edges. Based on the theory of intuitionistic fuzzy set (IFS) \cite{Ref38}\cite{Ref39}\cite{Ref40}\cite{Ref41}, a novel edge extraction and enhancement method has been put forward.

The IFS (denoted as $D$) of a given domain $X$ can be expressed as:
\begin{equation}
D=\left\{\left\langle x,\mu_D\left(x\right),\varphi_D\left(x\right)\right\rangle|x\in X\right\} 
\end{equation}
where $\mu_D:X\rightarrow\left[0,1\right]$ and $\varphi_D:X\rightarrow\left[0,1\right]$ are the membership function and the non-membership function of set D, respectively. Moreover, for all $x\in X$ in set $D$, we have $0\le\mu_D\left(x\right)+\varphi_D(x)\le1$.

In order to better describe the information of membership degree in set $D$, a hesitation function $\pi_D\left(x\right)$ has been employed, which combines both the membership and the non-membership functions. Also, for all $x\in X$ in set $D$, we have $0\le\pi_D\left(x\right)\le1$. Based on the functions above, the intuitionistic fuzzy entropy $E(D)$ of IFS $D$ in a given domain $X$ is defined as:
\begin{equation}
E\left(D\right)=\frac{1}{n}\sum_{i=1}^{n}\frac{2\mu_D\left(x_i\right)\varphi_D\left(x_i\right)+{\pi_D}^2(x_i)}{{\mu_D}^2\left(x_i\right)+{\varphi_D}^2\left(x_i\right)+{\pi_D}^2(x_i)} 
\end{equation}
where $n$ is the number of elements in set $D$.

Based on the gradient information of image, the local intuitionistic fuzzy entropy (LIFE) \cite{Ref42} of a pixel $u(i,j)$ in the image reflects the possibility of whither this pixel locates in the smooth regions. Therefore, the LIFE can be used to extract the edges in the image. The definition of functions $\mu_D(i,j)$, $\varphi_D(i,\ j)$ and $\pi_D\left(i,j\right)$ is:
\begin{equation}
\mu_D\left(i,j\right)={(1-\nabla u_{norm}\left(i,j\right))}^{\lambda(\lambda+1)}  
\end{equation}
\begin{equation}
\varphi_D\left(i,j\right)=1-{(1-\nabla u_{norm}\left(i,j\right))}^\lambda
\end{equation}
\begin{equation}
\pi_D\left(i,j\right)=1-\mu_D\left(i,j\right)-\nu_D(i,j)
\end{equation}
where ${\nabla u}_{norm}$ denotes the normalized gradient field of image. And the LIFE of pixel $u$ in its $n \times n$ neighborhood can be expressed as:
\begin{equation}
E\left(u\right)=\frac{1}{n\times n}\sum_{m=1}^{n\times n}\frac{2\mu_D\left(m\right)\varphi_D\left(m\right)+{\pi_D}^2(m)}{{\mu_D}^2\left(m\right)+{\varphi_D}^2\left(m\right)+{\pi_D}^2(m)}
\end{equation}
Usually, to maximum the LIFE, we have $\lambda=argmax{E\left(u\right)}$.

One advantage of LIFE in edge detection is that by simply changing the rule in calculating the gradient field, the directional LIFE under directional gradient field can be acquired to better extract the edges that are enhanced in the US image, since the edges parallel to the propagation direction of ultrasound seldom reflects ultrasound and are not enhanced in the US image.

In order to better extract the enhanced edges in the CT image, the radial directional gradient field under the polar coordinates, $\nabla u_r$, is put forward. In this method, the gradient vector ${(\nabla u}_x\left(i,j\right)$,$\ {\nabla u}_y(i,j))$ of a given pixel $(i, j)$ under the rectangular coordinate system is calculated firstly. Then, the deflection angle $\theta(i, j)$ is acquired using the coordinates of this pixel and the center of the sectorial view field of the US image, which is also the origin of the polar coordinate system. As shown in Fig. 2 and Equation (10).

\begin{equation}
\theta(i,j)=\arctan\left(\frac{{i-x}_0}{j-y_0}\right) 
\end{equation}

where $(x_0, y_0)$ is the coordinate under the rectangular coordinate system of the center of the circular-arc-shaped upper and lower boundaries of the ultrasonic image.
\begin{figure}
  \centering  
      \includegraphics[width=0.665\linewidth]{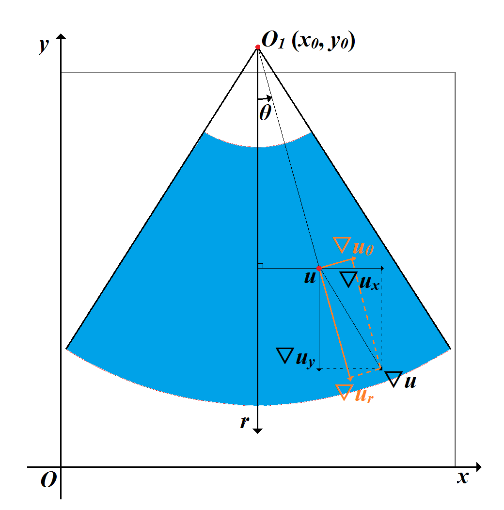}
      \caption{The 2D Cartesian coordinates and polar coordinates.}
 \label{fig:fig2}
\end{figure}

\begin{equation}
\nabla u_r\left(i,j\right)={\nabla u}_x\left(i,j\right)\sin{\left(\theta(i,j)\right)}+{\nabla u}_y\left(i,j\right)\cos\left(\theta\left(i,j\right)\right)
\end{equation}

After that, the $\nabla u_r(i,j)$ of pixel $(i, j)$ can be obtained by Equation (11). And using the normalized $\nabla u_r$, the radial directional LIFE, $E_r(i,j)$, can be acquired. In our experiences, in order to reduce the time used for calculation, $\lambda$ is set to be 4. Using a soft-threshold of 1.5 to reduce the weak textures, as shown in Equation (12a), where $sgn()$ is the sign function, as shown in Equation (12b). Then we can get the extracted edges. Finally, we add the extracted edges to the de-noised CT slice to enhance the slice. 
\begin{subequations}
\begin{equation}
E_r\left(i,j\right)=\left\{  
\begin{array}{ll}E_r\left(i,j\right)-sgn(E_r\left(i,j\right))\times1.5,\qquad \qquad \left|E_r\left(i,j\right)\right|\ge1.5\\
0,\qquad\qquad \qquad \qquad \qquad \qquad \qquad \qquad \; \left|E_r\left(i,j\right)\right|<1.5
\end{array}  
\right. 
\end{equation}%
\begin{equation}%
sgn\left(x\right)=\left\{  
\begin{array}{lll} 1,\quad x>0\\
0,\quad x=0\\
-1,\; x<0
\end{array}  
\right. 
\end{equation}
\end{subequations}

\subsection{The Non-rigid registration and resampling of the de-noised US image}
In order to reduce in influence of heart, bone or other organs that are not exist in the US image on the non-registration result, and acquire better performance of the registration algorithm on the liver, the region of liver is marked out using a mask on the enhanced CT slice. Then the liver in the US image is also extracted using the same mask.

After the extraction of the liver, the extracted regions in the CT slice and US image are firstly aligned by rigid translation. This is because that in the extraction of corresponding slice, the movement of other organs may influence the accuracy of registration on the liver, and the enhancement of edges can also improve the accuracy of rigid registration. After that, the 2D US to 2D CT non-rigid registration of the aligned regions is employed to acquire the deformation field. The Demons model based on the mutual information force is employed in this work to correct the non-rigid deformation of images in aligned regions, as shown in Section Appendix A.

For the region outside the mask, a Gaussian blur filter has been used to smooth the deformation field and avoid the sudden change on the boundaries of the mask.

Finally, the de-noised US image is resampled by the deformation field, and the resampled image is the output of our method. Then we compare the output US image with the extracted corresponding CT slice to evaluate our method.

\section{Results}
In this work, test on the registration between ultrasonic images and CT data of one patient form the Beijing Tsinghua Changgung Hospital has been done to evaluate our algorithm. In order to get a clearer view about the liver and blood vessels and get better registration performances, the window width and position of the CT data are adjusted before the registration.

\begin{figure}
  \centering  
      \includegraphics[width=0.965\linewidth]{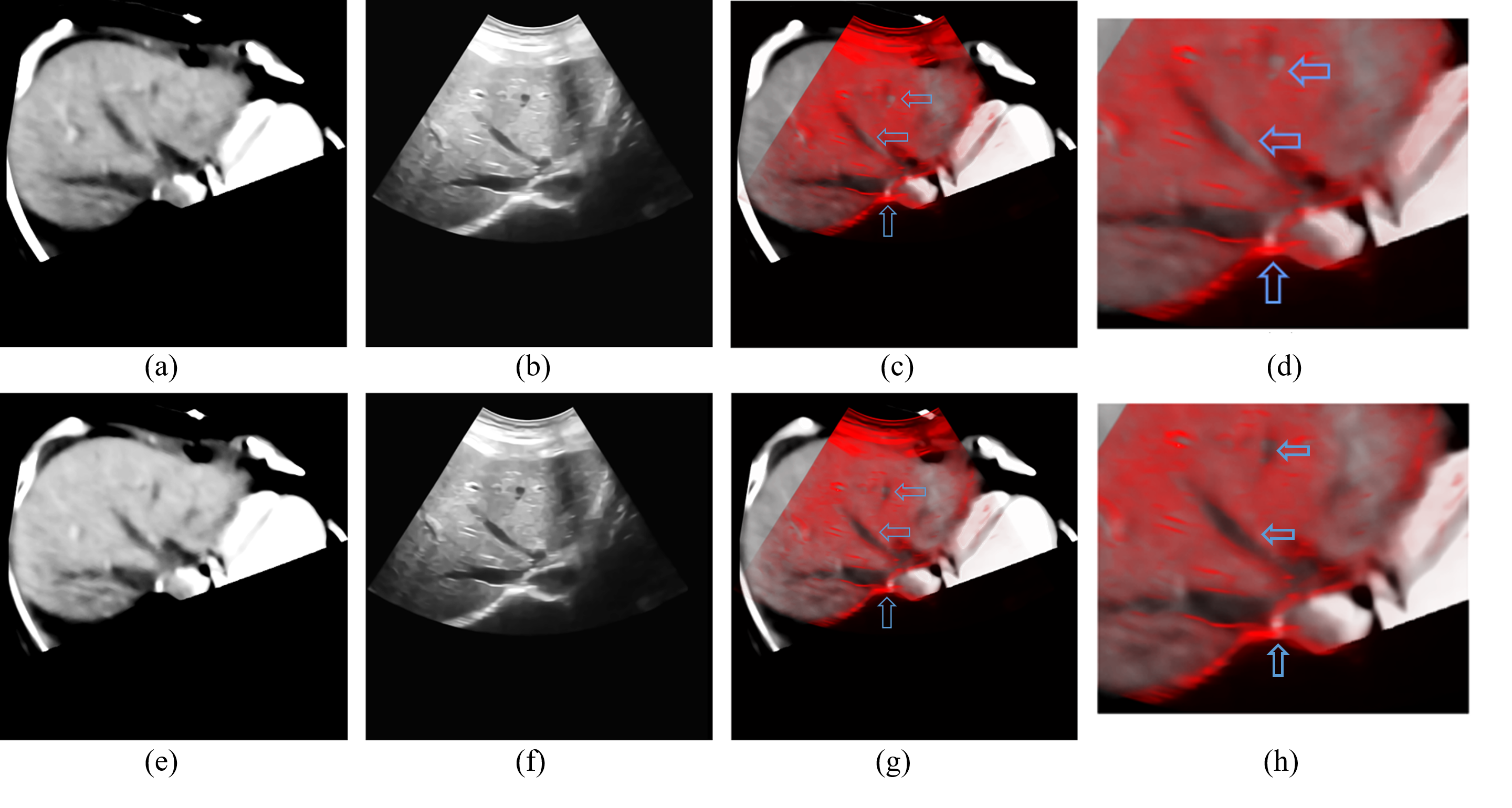}
      \caption{The data before and after registration of test 1. (a) is the corresponding CT slice of the original ultrasonic image (b) after the matching of world coordinate system. (e) is the result of the rigid registration from CT volume data to ultrasonic image. (f) is the non-rigid registration result of the ultrasonic image (b). (c) and (g) are the fusion image of (a) (b) and (e) (f), (d) and (h) are the details of (c) and (g), respectively.}
 \label{fig:fig3}
\end{figure}

The data before and after the registration are shown in Fig. 3-5, where subfigures 3(a), 4(a) and 5(a) are CT slices corresponding to original ultrasonic images, subfigures 3(b), 4(b) and 5(b) after the matching of the world coordinate system. Subfigures 3(e), 4(e) and 5(e) are CT slices after rigid registration to original ultrasonic images, and subfigures 3(f), 4(f) and 5(f) are non-rigid registration results of original ultrasonic images to their correspondence rigid registered CT slices. And fusion images of CT slices and ultrasonic images before and after registration are obtained and shown in Figs. 3(c) – 5(c) and 3(f) – 5(f) to better evaluate our algorithm. According to these figures, on comparison with the fusion image before registration, boundaries of both the liver and vessels are better matched. And the positions and shapes of vessels are also corrected. And the positions of bones in Figs. 4(g) and 5(g) are also closer to their positions in the ultrasonic images, which can be inferred by their “shadows” in the ultrasonic image.
 
\begin{figure}
  \centering
      \includegraphics[width=0.965\linewidth]{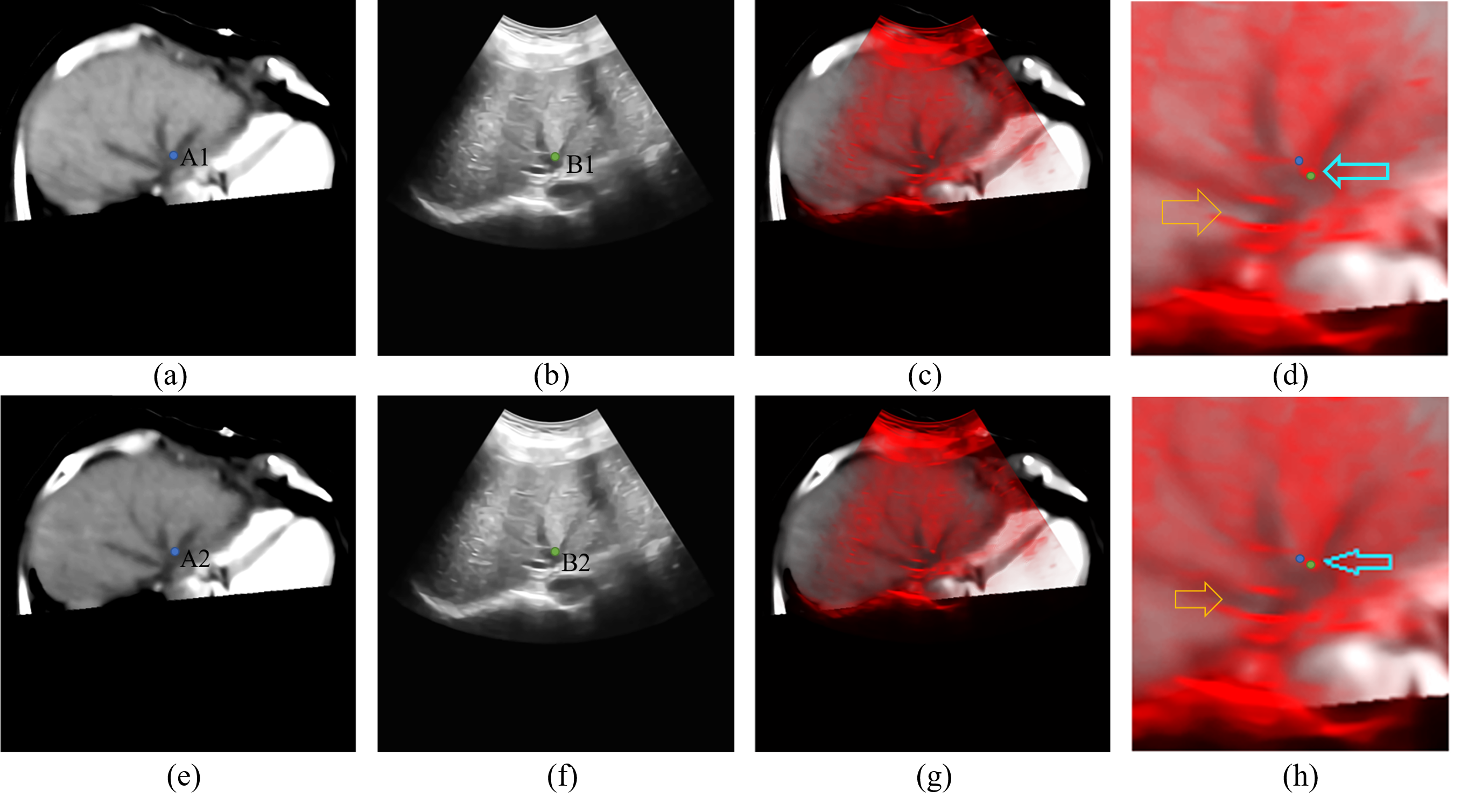}
      \caption{The data before and after registration of test 2. (a) is the corresponding CT slice of the original ultrasonic image (b) after the matching of world coordinate system. (e) is the result of the rigid registration from CT volume data to ultrasonic image. (f) is the non-rigid registration result of the ultrasonic image (b). (c) and (g) are the fusion image of (a) (b) and (e) (f), (d) and (h) are the details of (c) and (g), respectively.}
 \label{fig:fig4}
\end{figure}

\begin{figure}
  \centering
      \includegraphics[width=0.965\linewidth]{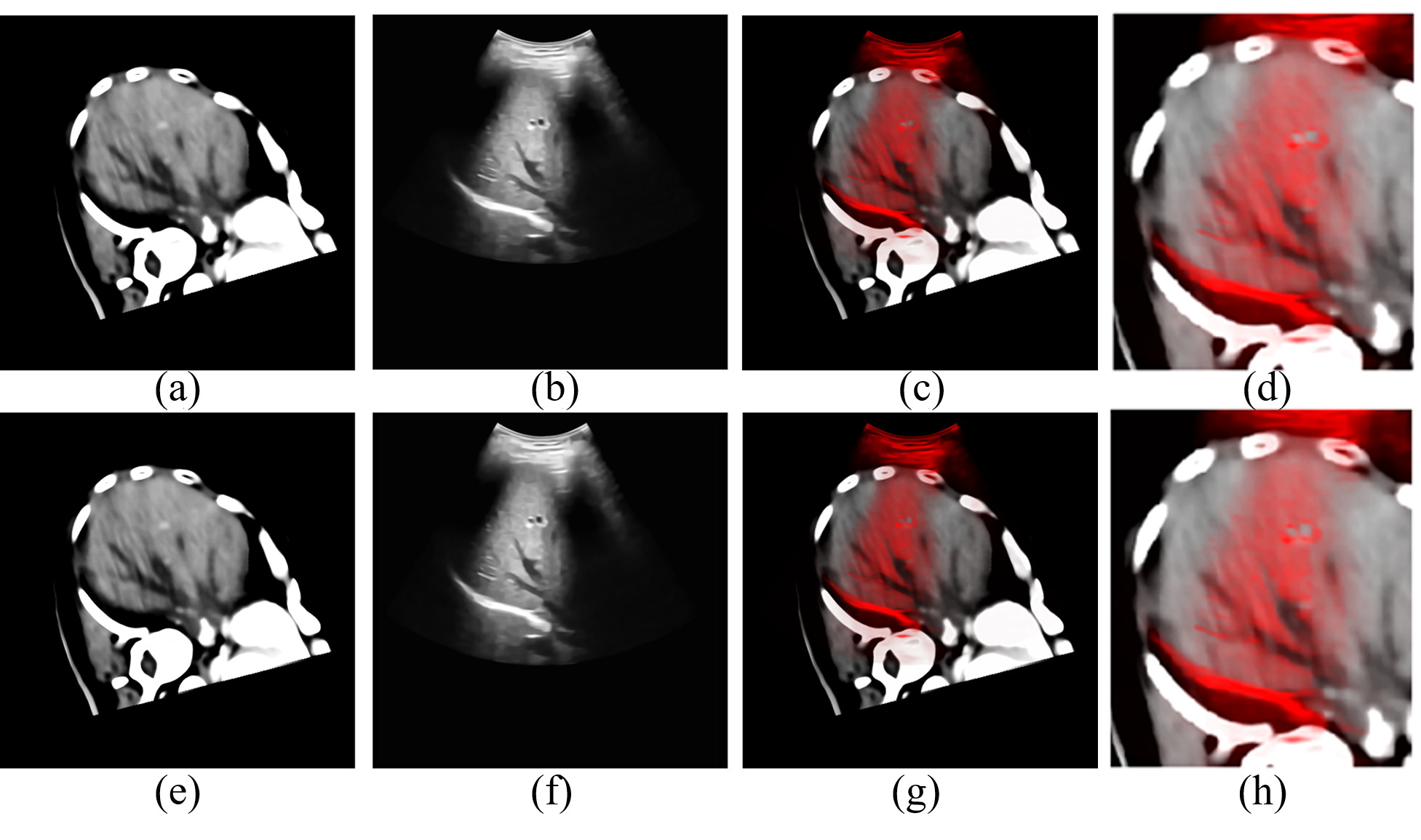}
      \caption{The data before and after registration of test 3. (a) is the corresponding CT slice of the original ultrasonic image (b) after the matching of world coordinate system. (e) is the result of the rigid registration from CT volume data to ultrasonic image. (f) is the non-rigid registration result of the ultrasonic image (b). (c) and (g) are the fusion image of (a) (b) and (e) (f), (d) and (h) are the details of (c) and (g), respectively.}
 \label{fig:fig5}
\end{figure}

The details of results in Figs. 3-5 are shown in subfigures (g) and (h) of Figs. 3-5, which also show that our registration algorithm can improve the matching accuracy between the CT data and ultrasonic images, especially in the parts pointed by arrows in Figs. 3 and 4. However, although our algorithm has better registration performances than the Demons algorithm, there still exists some parts that can be improved, such as the places pointed by the blue arrow in Fig. 4(h) although the distance between points A2 and B2 is shorter than that between points A1 and B1, points A2 and B2 are still not matched. 
\begin{figure}
  \centering
      \includegraphics[width=0.965\linewidth]{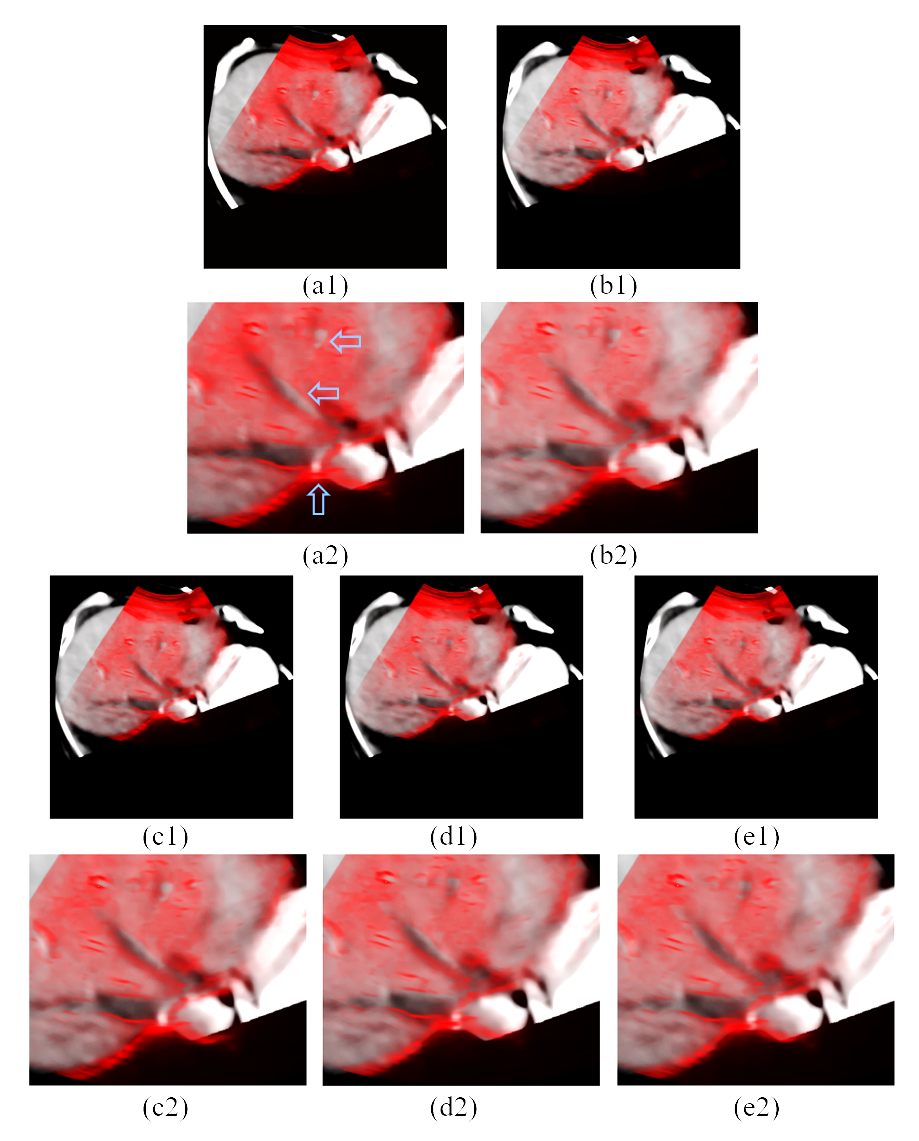}
      \caption{The comparison of registration accuracy of different methods using data in Fig. 3. (a1) is the corresponding CT slice of the original ultrasonic image, (b1) is the fusion image of the CT and ultrasonic images after rigid registration, and (c1), (d1), (e1) are the fusion image of CT slice and non-rigid registration results of ultrasonic image using the original Demons method, Demons method with the simulated ultrasonic image of the CT slice by Field II and our method, respectively. (a2) – (e2) are the details of (a1) - (e1), respectively.}
 \label{fig:fig6}
\end{figure}

\section{Discussion}
The registration results of Demons method and the Demons method uses the simulated ultrasonic image by Field II as well as our method on the data of Fig. 3 are compared, as shown in Fig. 4. Both the Field II method and our method can improve the matching accuracy on comparison with the rigid registered results. However, the original Demons method failed to improve the registration accuracy. The original Demons method enlarged the distance between the corresponding vessels and the liver boundaries. On comparison with the result of the Demons method, the registration accuracy is improved with the help of the Field II method. And our method further improved registration performance of either blood vessels or the boundaries of liver. The Demons method fails to register the liver and the vessels, since the reflection by boundaries of liver and vessels changes the “structure” of the image, and thus influences the mutual information in the non-rigid registration and leads to the wrong result. Field II is efficient in the improvement of the similarity between the real ultrasonic image and the simulated image, however, this method is weak in the simulation of the boundaries. 

To better compare the results, the details of Figs. 3 and 4 are shown in Fig. 5, and the gray level values of pixels in the position of the red, blue and green lines on Figs. 6 (a) and (b) are shown in Figs. 6(c) – (e). Figs. 5 and 6 further proves that our method can acquire better results, since the registered ultrasonic image better matches the reference CT image. And according to the line charts, curves of our method have better matching accuracy to the reference curve of the CT image, since these curves have the best matching accuracy in the positions and lengths of the peaks and valleys and the closest distances between the position of the corresponding boundaries of the peaks and valleys.

\begin{figure}
  \centering
      \includegraphics[width=0.965\linewidth]{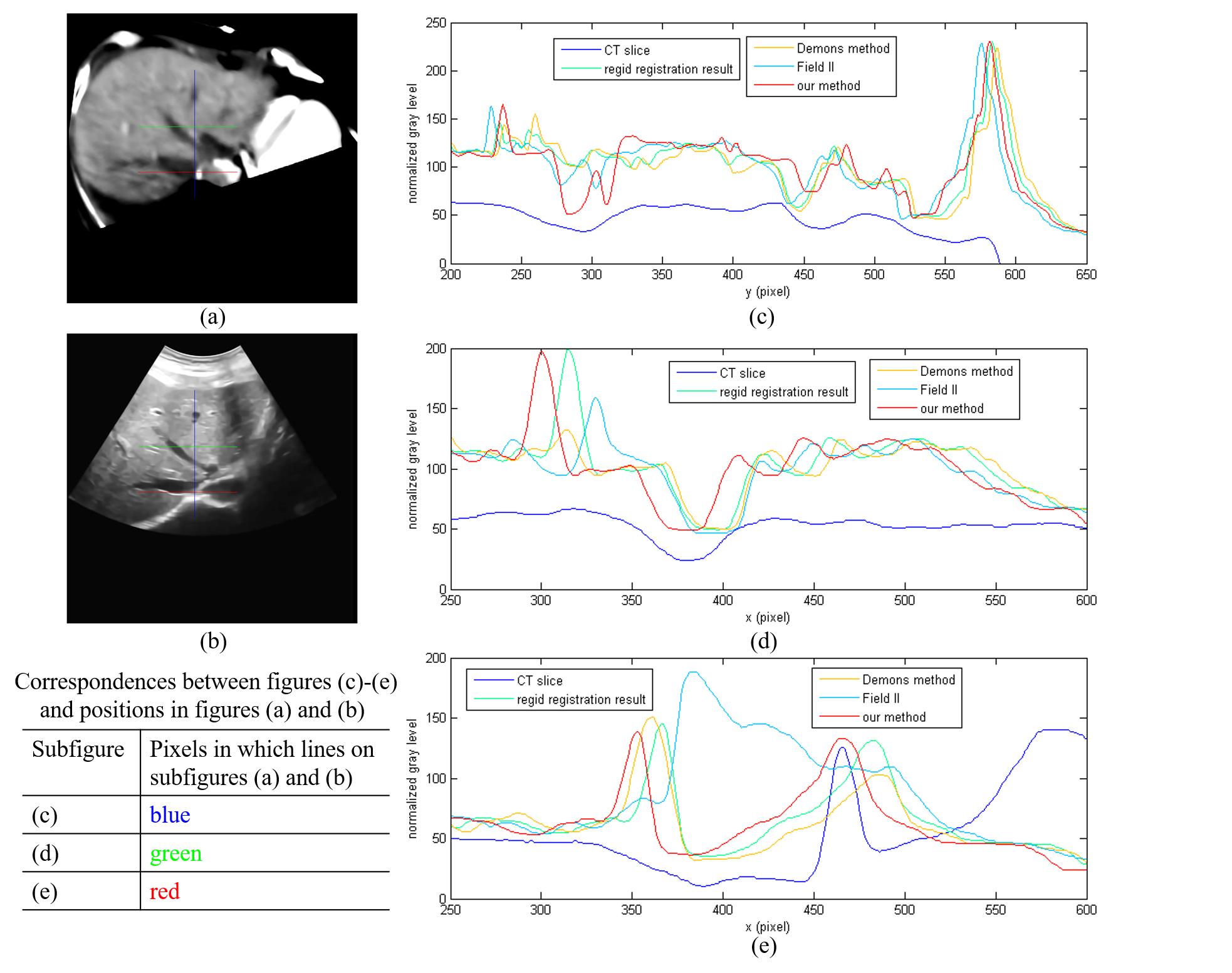}
      \caption{The gray level value distribution curves about pixels on the red, green and blue lines of the registration results of different methods. (a) and (b) show the position of these lines on the rigid-registered CT slice and the registration result of ultrasonic image of our method, respectively. (c) – (e) are curves about pixels on the blue, green and red lines.}
 \label{fig:fig7}
\end{figure}

Moreover, our method is less time-consuming than the Field II method, since it requires less calculation. It only takes about 2 seconds for the boundary enhancement approach in a $1024\ pixels \times 1024\ pixels$ CT slice on Matlab R2014a on a computer with 3.4 GHz and 3.41 GHz Intel i5-7500U CPU. And as for the Field II method, the calculation time is more than 10 hours. As for our method, the most time-consuming step, the BM4D denoising of CT slices, can be done before the ultrasonic examination, and the total operation time of our method to register one ultrasonic image is about 30 seconds. Therefore, our method can be used in the surgery.

Reducing the calculation time and extend the application of this algorithm to ultrasonic images that boundaries are not enhanced are our research directions in the future. With the development of deep learning algorithms, several ultrasonic image simulation algorithms based on deep learning has been put forward \cite{Ref43}, which may be useful in improving the registration accuracy and reducing the calculation time. And the research work on these algorithms will also be our future work.

\section{Conclusion}
The Demons method based on mutual information force and Quasi-Newton method is an efficient method in the registration of multi-model medical data. However, its performance needs to be improved when it comes to the registration between CT volume data and ultrasonic images. We put forward a novel boundary extraction and enhancement method based on radial directional local intuitionistic fuzzy entropy in the polar coordinates to improve the similarity between original ultrasonic images and enhanced CT slices with less calculation than existed ultrasonic image simulation methods. We furtherly designed a new workflow for the registration between CT volume data and ultrasonic images after the matching of the world coordinate system. Experiments shows that our method can acquire better registration accuracy than the original Demons method and the Demons method based on simulated images by Field II. The operation time of our edge detection and enhancement is about 2 seconds, and that of the whole workflow is about 30 seconds for the registration on $1024 \times 1024 \times 31$-pixel CT volume data and $1024 \times 1024$-pixel ultrasonic image, and it can be used in the surgery.

Future work mainly includes the improvement on the performance of large deformation correction and reducing the operation time using parallel computation methods. Using the deep learning methods to obtain simulated CT or ultrasonic data with higher similarity in shorter time may also be useful in increasing the registration accuracy. Extending the application of this method to the ultrasonic images whose boundaries are not enhanced is another research direction in the future. Moreover, more non-rigid registration algorithms will be studied in the future to improve the registration accuracy and reduce the calculation time, such as the iLogDemons method \cite{Ref44}, the Hybrid feature-based Diffeomorphic registration method \cite{Ref45} and deep learning-based registration methods \cite{Ref46}.

\section*{Acknowledgements}
The author thank the Beijing Tsinghua Changgung Hospital for providing the CT volume data and the ultrasonic images. The author also thank the Mr. Zan Liu, Ms. Wenjun Yu, Ms. Weijuan Li and other colleagues in Shanghai United-imaging Co., Ltd who have given the registration software of Demons method and the useful suggestions and discussions.
\section*{Funding}
This work was supported by the National Key R\&D Program of China (2017YFC0112801) and the Key Project of Special Development Foundation of Shanghai Zhangjiang National Innovation Demonstration Zone (1701-JD-D1112-030).
\section*{Disclosures}
The authors declare that there are no conflicts of interest related to this article.

 \appendix
  \renewcommand{\appendixname}{Appendix~\Alph{section}}
  \setcounter{table}{0}
  \setcounter{equation}{0}
  \renewcommand{\theequation}{\thesection.\arabic{equation}}
  \setcounter{figure}{0}
  \renewcommand{\thefigure}{\thesection.\arabic{figure}}
  \section*{Appendix}
  \section{The Demons algorithm based on the mutual information force}
In the registration of multi-modal images, the mutual information is widely used as the similarity measure. Mutual information (MI) is a basic conception in information theory. And it is usually used to describe the statistical correlations between two between two random variables or the amount of information that one variable contains about the other. The MI registration criterion presented here states that the MI of the image intensity values of corresponding voxel pairs is maximal if the images are geometrically aligned \cite{Ref47}. The mutual information between two systems $A$ and $B$, $I(A,B)$, is defined using the entropy, as shown in Equation (A.1).
\begin{equation}%
   I\left(A,B\right)=H\left(A\right)+H\left(B\right)-H\left(A,B\right)
\end{equation}
where $H\left(A\right)$ and $H\left(B\right)$ are the entropy of each system, and $H\left(A,B\right)$ is the joint entropy between these systems. As shown in Equations (A.2) and (A.3).
\begin{equation}%
   H\left(A\right)=-\sum_{a}{p_A\left(a\right)\log{p_A\left(a\right)}}
\end{equation}
\begin{equation}%
   H\left(A,B\right)=-\sum_{a,b}{p_{AB}\left(a,b\right)\log{p_{AB}\left(a,b\right)}}
\end{equation}

where $p_A\left(a\right)$ and $p_{AB}\left(a,b\right)$ are the marginal probability density function and the joint probability density function, respectively. Therefore, the mutual information $I\left(A,B\right)$ between images $A$ and $B$ can be calculated by Equation (A.4).
\begin{equation}%
   I\left(A,B\right)=\sum_{a,b}{p_{AB}\left(a,b\right)\log{\frac{p_{AB}\left(a,b\right)}{p_A\left(a\right)\bullet p_B\left(b\right)}}}
\end{equation}
\begin{equation}%
   NMI\left(A,B\right)=\frac{H\left(A\right)+H\left(B\right)}{H\left(A,B\right)}
\end{equation}

Moreover, the normalized measure of mutual information, $NMI\left(A,B\right)$, proposed by Studholme et al \cite{Ref48} is less sensitive to changes in overlap, as shown in Equation (A.5) \cite{Ref49}. Therefore, it is employed in the non-rigid registration as the similarity measure in this work.

The Demons model is based on the principle of intensity conservation between image frames, and the non-rigid registration can be regarded as a diffusion progress from the source image $B$ to the target image $A$. As shown in Equation (A.6) \cite{Ref50}\cite{Ref51}\cite{Ref52}\cite{Ref53}\cite{Ref54}\cite{Ref55}: 
\begin{equation}%
   \vec{v}\cdot \nabla i_b=i_a-i_b
\end{equation}
where $\vec{v}=(v_x,v_y)$ in the registration of 2D images, which donates the estimated displacement, or the velocity for a point $u_a$ with intensity $i_a$ in image $A$ to match the corresponding point $u_b$ with intensity $i_b$ in image $B$. $\vec{\nabla}i_b$ is the gradient of the target image $B$. The term ${h(i_a-i_b)}^2$ is used to stabilize the velocity equation.
\begin{equation}%
   \vec{v}=\frac{(i_a-i_b)\nabla i_b}{\left | \left | \nabla i_b\right | \right |^2 + {h(i_a-i_b)}^2}
\end{equation}
In the multi-modal images’ registration, the mutual information force $\vec{F}(i,j,t)$ is employed to calculate the velocity of pixel $(i,\ j)$ at time $t$. Equation (A.8) shows the mutual information force in the $x$-direction, $F_x$ \cite{Ref56}. 
\begin{equation}%
   v_x=F_x\left(i,j,t\right)=\frac{\partial I\left(A,B\left(t\right)\right)}{\partial u_x}=\frac{1}{n\times n}ln\left[\frac{p_{mr}}{P_r}/{\frac{p_{mt}}{P_t}}\right]  
\end{equation}

where $m$ is a square window in the target image $A$, and $r$ and $t$ denote pixels of the source image $B$ in the squire windows on the left side and right side of the corresponding window $s$ in image $B$ of window $m$ in image $A$. $p_{mr}$ and $p_{mt}$ are joint intensity probabilities between the specified windows, and $P_r$ and $P_t$ are marginal intensity probabilities for the source and target images, respectively.

Component of mutual information force in y-direction follows similarly where probabilities associated with neighboring voxels in $y$-direction will replace those in Equation (A.8).

The specific steps of Demons registration algorithm are as follows:

\begin{table}[h]
\caption{The specific steps of Demons registration algorithm}
\label{table_A1}
\centering
\begin{tabular}{l}
\hline
\textbf{Algorithm 1.} The Demons registration algorithm\\
\hline
		1. \textbf{Input:} two images $A$, $B$, Gaussian filter parameter $\sigma$, relaxation factor $\alpha$, maximum iteration number $max\_iter$ \\
		2. Initialize deformation field, $T_0$.\\
	    3. \textbf{While} $k<max\_iter$ and convergence not reached \textbf{do}\\
	    4. \quad Calculate $v_k$\\
	    5. \quad Regularize $v_k$ with a Gaussian filter. $v_k=G_\sigma\ast v_k$ \quad \quad (A.9)\\
	    6. \quad Calculate $\alpha$, and use it to insure a displacement smaller than a pixel\\
	    7. \quad Update $T_{k+1}=T_k+\alpha\times v_i$ \quad \quad (A.10)\\
	    8. \textbf{End while}\\
        9. \textbf{Output:} deformation field $T_k$.\\

\hline
\end{tabular}
\end{table}

  \section{The BM3D and BM4D method}
  
The BM3D method contains two main steps, which are basic estimation (step 1 in Fig. (B.1)) and final estimation (step 2 in Fig. (B.1)) \cite{Ref26}\cite{Ref27}\cite{Ref28}. Fig. (B.1) shows the flowchart of BM3D algorithm. The detailed steps are similar in each main step, which are grouping of the 2D blocks (images in each group form a 3D array), calculating local estimates by the three-dimensional collaborative filtering in the spectrum domain (3D hard-threshold method in step1, and Wiener filtering in step 2) of each 3D array and the aggregation of the weighted means of the local estimates. The preliminarily de-noised basic estimate results are employed in the step 2 to improve the accuracy of grouping and to acquire more accurate filtering results by using it as the pilot signal of the empirical Wiener filtering. 

\begin{figure}
  \centering
      \includegraphics[width=0.965\linewidth]{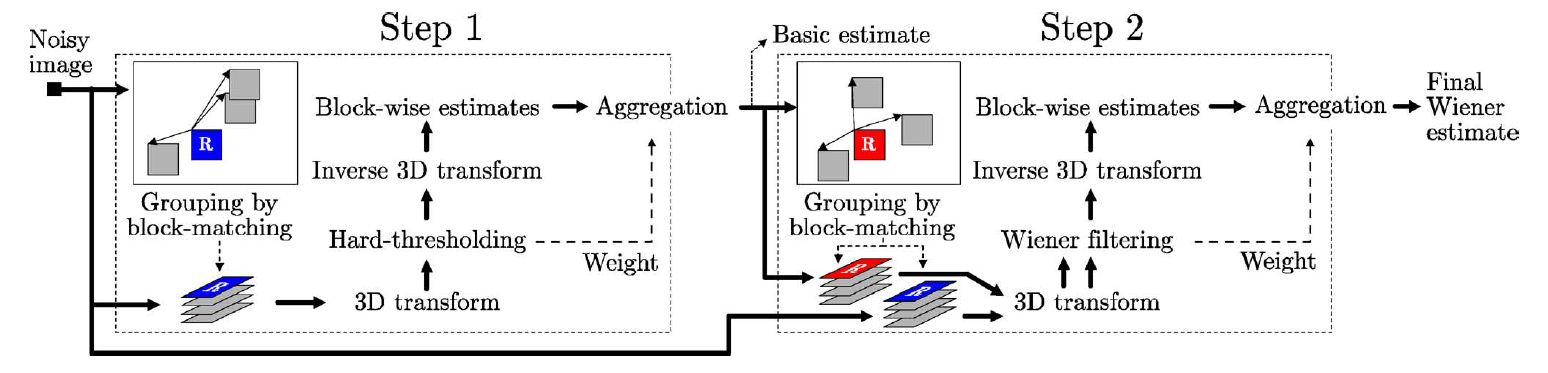}
      \caption{The flowchart of BM3D method. Fig. B.1 is from Reference \cite{Ref26}.}
  \label{fig:fig2_1}
\end{figure}
The BM4D algorithm is an extension of BM3D algorithm into 3D volumetric data. The flowchart of BM4D algorithm is shown in Fig. B.2. It has the similar principle and processes as the BM3D method, but uses cubes of voxels instead of the blocks of pixels as basic data patches to form the 4D group \cite{Ref29}. It has good de-noising performance, but it is more time-consuming. Therefore, before the registration, the CT volume data is de-noised firstly.
 
\begin{figure}
  \centering
      \includegraphics[width=0.965\linewidth]{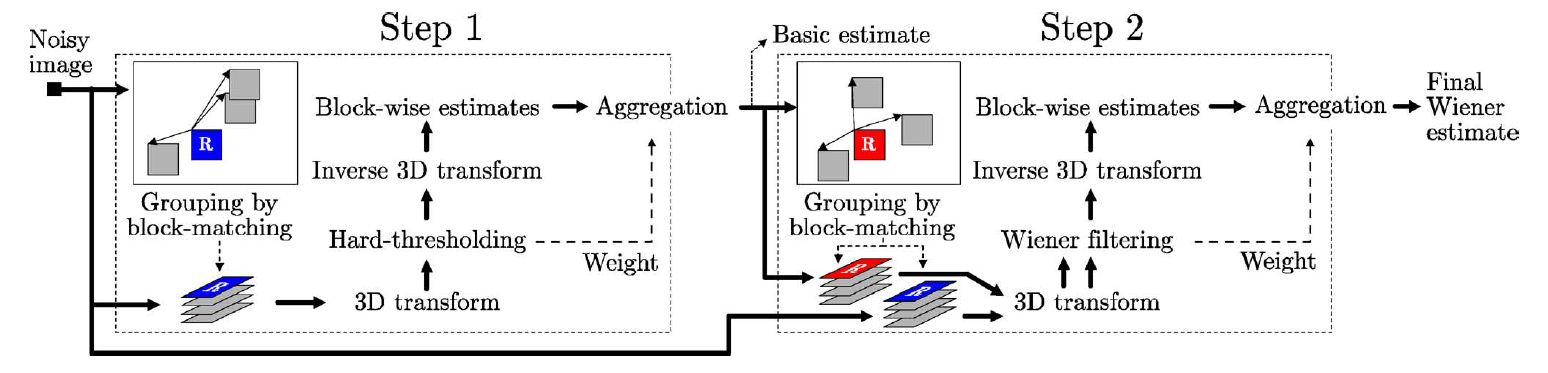}
      \caption{The flowchart of BM4D method. Fig. B.2 is from Reference \cite{Ref29}.}
  \label{fig:fig2_1}
\end{figure}


\bibliographystyle{unsrt}  
\bibliography{references}  

\begin{thebibliography}{10}

\bibitem{Ref1}
Haydar Talib, Matthias Peterhans, Jaime Garcia, Martin Styner, and Miguel
  Angel~Gonzalez Ballester.
\newblock Information filtering for ultrasound-based real-time registration.
\newblock {\em IEEE Transactions on Biomedical Engineering}, 58(3):531--540,
  2011.

\bibitem{Ref2}
Caroline Ewertsen, Adrian Săftoiu, Lucian Gruionu, S~Karstrup, and Michael~B
  Nielsen.
\newblock Real-time image fusion involving diagnostic ultrasound.
\newblock {\em American Journal of Roentgenology}, 200(3):W249--W255, 2013.

\bibitem{Ref3}
Kasper Marstal, Floris~F Berendsen, Marius Staring, and Stefan Klein.
\newblock Simpleelastix: A user-friendly, multi-lingual library for medical
  image registration.
\newblock In {\em Proceedings of the IEEE Conference on Computer Vision and
  Pattern Recognition Workshops}, pages 574--582. IEEE, 2016.

\bibitem{Ref4}
Erika R~E Denton, Luke~I Sonoda, Daniel Rueckert, Sheila Rankin, Carmel Hayes,
  Martin~O Leach, D~Hill, and David~J Hawkes.
\newblock Comparison and evaluation of rigid, affine, and nonrigid registration
  of breast mr images.
\newblock {\em Journal of Computer Assisted Tomography}, 23(5):800--805, 1999.

\bibitem{Ref5}
Aristeidis Sotiras, Christos Davatzikos, and Nikos Paragios.
\newblock Deformable medical image registration: A survey.
\newblock {\em IEEE Transactions on Medical Imaging}, 32(7):1153--1190, 2013.

\bibitem{Ref6}
Tim Mcinerney and Demetri Terzopoulos.
\newblock Deformable models in medical image analysis: a survey.
\newblock {\em Medical Image Analysis}, 1(2):91--108, 1996.

\bibitem{Ref7}
Tim Mcinerney and Demetri Terzopoulos.
\newblock Deformable models in medical image analysis.
\newblock In {\em Proceedings of the workshop on mathematical methods in
  biomedical image analysis}, pages 171--180. IEEE, 2002.

\bibitem{Ref8}
Chengqian Che, Tejas~Sudharshan Mathai, and John Galeotti.
\newblock Ultrasound registration: A review.
\newblock {\em Methods}, 115:128--143, 2017.

\bibitem{Ref9}
Terry~K Koo and Wingchi~E Kwok.
\newblock Hierarchical ct to ultrasound registration of the lumbar spine: A
  comparison with other registration methods.
\newblock {\em Annals of Biomedical Engineering}, 44(10):2887--2900, 2016.

\bibitem{Ref10}
Philip Pratt, Alexander Jaeger, Archie Hugheshallett, Erik Mayer, Justin Vale,
  Ara Darzi, Terry~M Peters, and Guangzhong Yang.
\newblock Robust ultrasound probe tracking: initial clinical experiences during
  robot-assisted partial nephrectomy.
\newblock {\em computer assisted radiology and surgery}, 10(12):1905--1913,
  2015.

\bibitem{Ref11}
Tony~C Poon and Robert Rohling.
\newblock Tracking a 3-d ultrasound probe with constantly visible fiducials.
\newblock {\em Ultrasound in Medicine and Biology}, 33(1):152--157, 2007.

\bibitem{Ref12}
Johann Hummel, Michael~R Bax, Michael Figl, Yan Kang, Calvin~R Maurer, Wolfgang
  Birkfellner, Helmar Bergmann, and Ramin Shahidi.
\newblock Design and application of an assessment protocol for electromagnetic
  tracking systems.
\newblock {\em Medical Physics}, 32(7):2371--2379, 2005.

\bibitem{Ref13}
Johann Hummel, Michael Figl, Michael~R Bax, Helmar Bergmann, and Wolfgang
  Birkfellner.
\newblock 2d/3d registration of endoscopic ultrasound to ct volume data.
\newblock {\em Physics in Medicine and Biology}, 53(16):4303--4316, 2008.

\bibitem{Ref14}
Roch~M Comeau, Abbas~F Sadikot, Aaron Fenster, and Terry~M Peters.
\newblock Intraoperative ultrasound for guidance and tissue shift correction in
  image‐guided neurosurgery.
\newblock {\em Medical Physics}, 27(4):787--800, 2000.

\bibitem{Ref15}
R~W Prager, Robert Rohling, A~H Gee, and L~H Berman.
\newblock Rapid calibration for 3-d freehand ultrasound.
\newblock {\em Ultrasound in Medicine and Biology}, 24(6):855--869, 1998.

\bibitem{Ref16}
Andrew Fitzgibbon.
\newblock Robust registration of 2d and 3d point sets.
\newblock {\em Image and Vision Computing}, 21(13):1145--1153, 2003.

\bibitem{Ref17}
Parastoo Farnia, Alireza Ahmadian, Alireza Khoshnevisan, Amirhossein
  Jaberzadeh, Nasim~Dadashi Serej, and Anahita~Fathi Kazerooni.
\newblock An efficient point based registration of intra-operative ultrasound
  images with mr images for computation of brain shift; a phantom study.
\newblock In {\em 2011 Annual International Conference of the IEEE Engineering
  in Medicine and Biology Society}, pages 8074--8077. IEEE, 2011.

\bibitem{Ref18}
Diego D~B Carvalho, Stefan Klein, Zeynettin Akkus, Anouk~C Van~Dijk, Hui Tang,
  Mariana Selwaness, Arend F~L Schinkel, Johan~G Bosch, Aad~Van Der~Lugt, and
  Wiro~J Niessen.
\newblock Joint intensity‐and‐point based registration of free‐hand
  b‐mode ultrasound and mri of the carotid artery.
\newblock {\em Medical Physics}, 41(5):052904, 2014.

\bibitem{Ref19}
Hang Gao, Torbjorn Hergum, Hans Torp, and Jan Dhooge.
\newblock Comparison of the performance of different tools for fast simulation
  of ultrasound data.
\newblock {\em Ultrasonics}, 52(5):573--577, 2012.

\bibitem{Ref20}
Wolfgang Wein, Shelby Brunke, Ali Khamene, Matthew~R Callstrom, and Nassir
  Navab.
\newblock Automatic ct-ultrasound registration for diagnostic imaging and
  image-guided intervention.
\newblock {\em Medical Image Analysis}, 12(5):577--585, 2008.

\bibitem{Ref21}
Haoran Jin, Ruochong Zhang, Siyu Liu, and Yuanjin Zheng.
\newblock Fast and high-resolution three-dimensional hybrid-domain
  photoacoustic imaging incorporating analytical-focused transducer beam
  amplitude.
\newblock {\em IEEE Transactions on Medical Imaging}, 38(12):2926--2936, 2019.

\bibitem{Ref22}
David Baek, J~Jensen, and Morten Willatzen.
\newblock Calibration of field ii using a convex ultrasound transducer.
\newblock {\em Physics Procedia}, 3(1):995--1001, 2010.

\bibitem{Ref23}
Jørgen~Arendt Jensen, Jrgen~Arendt Jensen, Svetoslav~Ivanov Nikolov, and
  Svetoslav~Ivanov Nikolov.
\newblock Fast simulation of ultrasound images.
\newblock In {\em 2000 IEEE Ultrasonics Symposium. Proceedings. An
  International Symposium (Cat. No. 00CH37121)}, pages 1721--1724. IEEE, 2000.

\bibitem{Ref24}
J~Jensen and N~B Svendsen.
\newblock Calculation of pressure fields from arbitrarily shaped, apodized, and
  excited ultrasound transducers.
\newblock {\em IEEE Transactions on Ultrasonics Ferroelectrics and Frequency
  Control}, 39(2):262--267, 1992.

\bibitem{Ref25}
Wolfgang Wein, Ali Khamene, Dirkandre Clevert, Oliver Kutter, and Nassir Navab.
\newblock Simulation and fully automatic multimodal registration of medical
  ultrasound.
\newblock In {\em International Conference on Medical Image Computing and
  Computer-Assisted Intervention}, volume~10, pages 136--143. Springer, 2007.

\bibitem{Ref26}
Kostadin Dabov, Alessandro Foi, Vladimir Katkovnik, and Karen Egiazarian.
\newblock Image denoising by sparse 3-d transform-domain collaborative
  filtering.
\newblock {\em IEEE Transactions on Image Processing}, 16(8):2080--2095, 2007.

\bibitem{Ref27}
Yingkun Hou, Chunxia Zhao, Deyun Yang, and Yong Cheng.
\newblock Comments on "image denoising by sparse 3-d transform-domain
  collaborative filtering".
\newblock {\em IEEE Transactions on Image Processing}, 20(1):268--270, 2011.

\bibitem{Ref28}
Marc Lebrun.
\newblock An analysis and implementation of the bm3d image denoising method.
\newblock {\em Image Processing On Line}, 2:175--213, 2012.

\bibitem{Ref29}
Matteo Maggioni, Vladimir Katkovnik, Karen Egiazarian, and Alessandro Foi.
\newblock Nonlocal transform-domain filter for volumetric data denoising and
  reconstruction.
\newblock {\em IEEE Transactions on Image Processing}, 22(1):119--133, 2013.

\bibitem{Ref30}
Vladimir Katkovnik, Alessandro Foi, Karen Egiazarian, and J~Astola.
\newblock From local kernel to nonlocal multiple-model image denoising.
\newblock {\em International Journal of Computer Vision}, 86(1):1--32, 2010.

\bibitem{Ref31}
Shuo Huang, Ping Zhou, Hao Shi, Yu~Sun, and Suiren Wan.
\newblock Image speckle noise denoising by a multi-layer fusion enhancement
  method based on block matching and 3d filtering.
\newblock {\em The Imaging Science Journal}, 67(4):224--235, 2019.

\bibitem{Ref32}
Shuo Huang, Le~Cheng, Bin Zhu, Ping Zhou, Yu~Sun, Bing Zhang, and Suiren Wan.
\newblock The time-domain integration method of digital subtraction angiography
  images.
\newblock {\em Computational and Mathematical Methods in Medicine},
  2018:5284969, 2018.

\bibitem{Ref33}
Suiren Wan, Balasundar~Iyyavu Raju, and Mandayam~A Srinivasan.
\newblock Robust deconvolution of high-frequency ultrasound images using
  higher-order spectral analysis and wavelets.
\newblock {\em IEEE Transactions on Ultrasonics Ferroelectrics and Frequency
  Control}, 50(10):1286--1295, 2003.

\bibitem{Ref34}
Jongsen Lee.
\newblock Digital image enhancement and noise filtering by use of local
  statistics.
\newblock {\em IEEE Transactions on Pattern Analysis and Machine Intelligence},
  2(2):165--168, 1980.

\bibitem{Ref35}
Shuji Jimbo and Akira Maruoka.
\newblock Expanders obtained from affine transformations.
\newblock In {\em Proceedings of the seventeenth annual ACM symposium on Theory
  of computing}, pages 88--97. ACM, 1985.

\bibitem{Ref36}
F~Lamare, T~Cresson, J~Savean, C~Cheze~Le Rest, A~J Reader, and D~Visvikis.
\newblock Respiratory motion correction for pet oncology applications using
  affine transformation of list mode data.
\newblock {\em Physics in Medicine \& Biology}, 52(1):121--140, 2007.

\bibitem{Ref37}
Erika R.~E. Denton, Luke~I. Sonoda, Daniel Rueckert, Sheila~C. Rankin, and
  David~J. Hawkes.
\newblock Comparison and evaluation of rigid, affine, and nonrigid registration
  of breast mr images.
\newblock {\em Journal of Computer Assisted Tomography}, 23(5):800--805, 1999.

\bibitem{Ref38}
L.~A. Zadeh.
\newblock Fuzzy sets.
\newblock {\em Information \& Control}, 8(3):338--353, 1965.

\bibitem{Ref39}
Krassimir~T Atanassov.
\newblock Intuitionistic fuzzy sets.
\newblock {\em Fuzzy Sets and Systems}, 20(1):87--96, 1986.

\bibitem{Ref40}
A~Gallegos Saliner, I.~Tsakovska, M.~Pavan, G.~Patlewicz, and A~P Worth.
\newblock The role of entropy in intuitionistic fuzzy contrast enhancement.
\newblock In {\em International Fuzzy Systems Association World Congress},
  pages 104--113. Springer, 2007.

\bibitem{Ref41}
Tamalika Chaira.
\newblock A rank ordered filter for medical image edge enhancement and
  detection using intuitionistic fuzzy set.
\newblock {\em Applied Soft Computing}, 12(4):1259--1266, 2012.

\bibitem{Ref42}
Yan Wang, Zhiguo Gui, Quan Zhang, and Yi~Liu.
\newblock Image denoising of pde based on local intuitionistic fuzzy entropy.
\newblock {\em Computer Engineering and Design}, 34(12):4256--4260, 2013.

\bibitem{Ref43}
Sanketh Vedula, Ortal Senouf, Alexander~M Bronstein, Oleg~V Michailovich, and
  Michael Zibulevsky.
\newblock Towards ct-quality ultrasound imaging using deep learning.
\newblock {\em arXiv preprint arXiv:1710.06304}, 2017.

\bibitem{Ref44}
Tommaso Mansi, Xavier Pennec, Maxime Sermesant, Herve Delingette, and Nicholas
  Ayache.
\newblock ilogdemons: A demons-based registration algorithm for tracking
  incompressible elastic biological tissues.
\newblock {\em International Journal of Computer Vision}, 92(1):92--111, 2011.

\bibitem{Ref45}
Amalia Cifor, Laurent Risser, Daniel Chung, Ewan~M Anderson, and Julia~A
  Schnabel.
\newblock Hybrid feature-based diffeomorphic registration for tumor tracking in
  2-d liver ultrasound images.
\newblock {\em IEEE Transactions on Medical Imaging}, 32(9):1647--1656, 2013.

\bibitem{Ref46}
Geert J~S Litjens, Thijs Kooi, Babak~Ehteshami Bejnordi, Arnaud A~A Setio,
  Francesco Ciompi, Mohsen Ghafoorian, Jeroen A W M~Van Der~Laak, Bram
  Van~Ginneken, and Clara~I Sanchez.
\newblock A survey on deep learning in medical image analysis.
\newblock {\em Medical Image Analysis}, 42:60--88, 2017.

\bibitem{Ref47}
Frederik Maes, A~Collignon, Dirk Vandermeulen, Guy Marchal, and Paul Suetens.
\newblock Multimodality image registration by maximization of mutual
  information.
\newblock {\em IEEE Transactions on Medical Imaging}, 16(2):187--198, 1997.

\bibitem{Ref48}
Colin Studholme, D~Hill, and David~J Hawkes.
\newblock An overlap invariant entropy measure of 3d medical image alignment.
\newblock {\em Pattern Recognition}, 32(1):71--86, 1999.

\bibitem{Ref49}
Josien P~W Pluim, J~B~A Maintz, and Max~A Viergever.
\newblock Mutual-information-based registration of medical images: a survey.
\newblock {\em IEEE Transactions on Medical Imaging}, 22(8):986--1004, 2003.

\bibitem{Ref50}
Jeanphilippe Thirion.
\newblock Image matching as a diffusion process: an analogy with maxwell's
  demons.
\newblock {\em Medical Image Analysis}, 2(3):243--260, 1998.

\bibitem{Ref51}
Dirk~Jan Kroon and Cornelis~H. Slump.
\newblock Mri modalitiy transformation in demon registration.
\newblock In {\em 2009 IEEE International Symposium on Biomedical Imaging: From
  Nano to Macro}, pages 963--966. IEEE, 2009.

\bibitem{Ref52}
Kinda~Anna Saddi and Farida Cheriet.
\newblock Large deformation registration of contrast-enhanced images with
  volume-preserving constraint.
\newblock In {\em Medical Imaging 2007: Image Processing. International Society
  for Optics and Photonics}, volume 6512, page 651203. SPIE, 2007.

\bibitem{Ref53}
Christophe Chefdhotel, Gerardo Hermosillo, and Olivier Faugeras.
\newblock Flows of diffeomorphisms for multimodal image registration.
\newblock In {\em IEEE International Symposium on Biomedical Imaging}, pages
  753--756. IEEE, 2002.

\bibitem{Ref54}
Torsten Rohlfing, Calvin~R Maurer, David~A Bluemke, and Michael~A Jacobs.
\newblock Volume-preserving nonrigid registration of mr breast images using
  free-form deformation with an incompressibility constraint.
\newblock {\em IEEE Transactions on Medical Imaging}, 22(6):730--741, 2003.

\bibitem{Ref55}
Hui Liu, Yunfei Yang, Zhenhong Shang, and Runxin Li.
\newblock Measurement of transverse velocity field of nvst solar high
  resolution image.
\newblock {\em Astronomical Research and Technology}, 15(2):151--157, 2018.

\bibitem{Ref56}
Gerardo Hermosillo, Christophe Chefd'Hotel, and Olivier Faugeras.
\newblock A variational approach to multi-modal image matching.
\newblock In {\em IEEE Workshop on Variational and Level Set Methods in
  Computer Vision}, pages 21--28. IEEE, 2001.

\end{thebibliography}






\end{document}